\documentclass[preprint]{revtex4}

\usepackage{graphicx}

\begin{document}

\title{Half-metallic Surface States and Topological Superconductivity
  in NaCoO$_2$ }

%\title{Topological superconductivity and Majorana fermions in
%  NaCoO$_2$/superconductor heterostructure}

\author{ Hongming Weng$^1$, Gang Xu$^1$, Haijun Zhang$^2$,
  Shou-Cheng Zhang$^{2,3}$, Xi Dai$^1$}

\email{daix@aphy.iphy.ac.cn}

\author{Zhong Fang$^1$}

\email{zfang@aphy.iphy.ac.cn}

\affiliation{$^1$ Beijing National Laboratory for Condensed Matter
  Physics, and Institute of Physics, Chinese Academy of Sciences,
  Beijing 100190, China;}

\affiliation{$^2$ Department of Physics, McCullough Building, Stanford
  University, Stanford, CA 94305-4045;}

\affiliation{$^3$ Center for Advanced Study, Tsinghua
  University,Beijing, 100084, China}

\date{\today}

\begin{abstract}
%    Majorana fermions obey non-Abelian statistics, which is the key
%    ingredient for the fault-tolerant topological quantum
%    computation. Recently, they have been proposed to exist in various
%    condensed-matter systems as emergent excitations. However,
%    experimental realization in these materials remains
%    challenging. For a half-metal, which acts as conductor for one
%    spin and insulator for another spin, the proximity effect with a
%    $s$-wave superconductor can induce topological $p_x+ip_y$
%    superconductivity and Majorana fermions in the vortex core. In
%    this work 
  Based on the first-principles calculations, we predict a novel
  half-metallic surface state in layered bulk insulator NaCoO$_2$,
  with tunable surface hole concentration.  The half-metallic surface
  has a single fermi surface with a helical spin texture, similar to
  the surface state of topological insulators, but with the key
  difference of time reversal symmetry breaking in the present
  case. We propose realization of topological superconductivity and
  Majorana fermions when the half-metallic surface states are in
  proximity contact with a conventional superconductor.
\end{abstract}


\maketitle

The possibility of realizing Majorana fermions in condensed matters as
emergent excitations has stimulated great current interest. Majorana
fermions are particles which are their own
antiparticles~\cite{majorana}. They constitute only half of a usual
fermion, and obey the non-Abelian statistics~\cite{non-Abelian}, which
is the key ingredient for the fault-tolerant topological quantum
computation~\cite{TQC}. The Majorana states associated with zero
energy mode have been predicted to exist in various
systems~\cite{non-Abelian,QH,p+ip-1,p+ip-2,proximity,Sarma,Alicea,Lee1,Lee2,Duck},
nevertheless, their experimental realizations remain challenging due
to the requirements of extreme conditions such as strong magnetic
field, low temperature, or ultra-clean samples. In this letter, we
will show that NaCoO$_2$/superconductor hetorestructure is a new and
simple platform for realizing such exotic states.

The simplest Majorana bound state is associated with a vortex core or
edge in a two-dimensional (2D) topological $p_x+ip_y$
superconductor~\cite{p+ip-1,p+ip-2}, which has a full pairing gap in
the bulk but with gapless chiral edge states (which consists of
Majorana fermions).  Unfortunately, such topological superconductors
are very rare in nature, this leads to the proposals of possible
``induced'' $p_x+ip_y$ order parameter through the proximity effect,
particularly for materials in contact with the simplest $s$-wave
superconductors~\cite{proximity,Sarma,Alicea,Lee1,Lee2,Duck}. The
proposal by Fu and Kane~\cite{proximity} is to use the surface state
of topological
insulators~\cite{Kane_PRL_2005,Bernevig_PRL_2006,Zhang}, where the
spin degeneracy is removed and yet strong proximity effect can be
expected due to the characteristic helical spin texture of Dirac-type
surface state. The experimental setting can be in principles obtained
from laboratory, however, since most known three-dimentional
topological insulators~\cite{Zhang,Hsieh,Xia,Chen} up to now are not
good bulk insulators and important surface states may overlap with
bulk states, experiments have to wait for development of
well-controlled clean samples. It was proposed recently that
semiconductor quantum wells with Rashba type SOC in proximity to
$s$-wave superconductor will produce similar
effect~\cite{Sarma,Alicea,Lee1,Lee2}. This may lower the experimental
threshold, since well-controlled samples are available nowadays. Both
proposals are encouraging, while experimental obstacles still
remain. First, and most importantly, magnetic insulating layers or
strong external magnetic field are required to break the time reversal
symmetry, which is not easy to implement experimentally; second, the
Fermi surfaces in both cases are too small, and fine control of
chemical potential is difficult for semiconductors in contact with a
superconductor.

To avoid the complication required by breaking time reversal symmetry
(such as the usage of magnetic insulating layers or magnetic field),
the most natural way is to start from a magnetic compound, rather than
non-magnetic topological insulators or semiconductor quantum wells. On
the other hand, however, the following ingredients have to be
satisfied over a wide energy or doping regime, in order to induce the
$p_x+ip_y$ superconductivity through proximity effect: (1) two
dimensional metal with a single (or an odd number of) Fermi surface;
(2) strong enough helical spin texture arising from spin-orbit
coupling (SOC). Those conditions all together suggest that a 2D single
band half-metal materials with strong enough SOC will be the best
candidate~\cite{Duck,HJ}. Following this strategy, we propose in this
work that NaCoO$_2$ is such a unique compound, which satisfies all
those conditions simultaneously.

\begin{figure}[tbp]
\includegraphics[clip,scale=0.25]{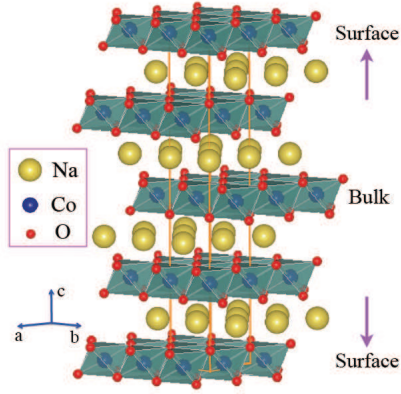}
\caption{(Color online) The structure of NaCoO$_2$ slab consisting of
  5 unit layers. After structure optimization, the surface Co-O
  bond-lengths are slightly modified (by 0.07\AA).}
\end{figure}

%\section{Results}

Bulk Na$_x$CoO$_2$ is a well known layered compound crystallizing in
planar triangle lattice with Co atom coordinated by oxygen
octahedrally (Fig.1). The Na atoms are interpolated between CoO$_2$
layers, and its concentration $x$ can be systematically tuned,
resulting in complex magnetic and electronic phase
diagram~\cite{NaCoO-phase}. In particular, the unconventional
superconductivity around $x$=0.35~\cite{NaCoO-SC}, and the layered
antiferromagnetic (AF) phase around
$x>0.65$~\cite{NaCoO-AF,NaCoO-neutron,NaCoO-neutron-2}, have drawn
lots of attentions. For our purpose, however, we consider the
stoichiometry NaCoO$_2$ (i.e. $x$=1.0) and its surface state.

\begin{figure}[tbp]
\includegraphics[clip,scale=0.5]{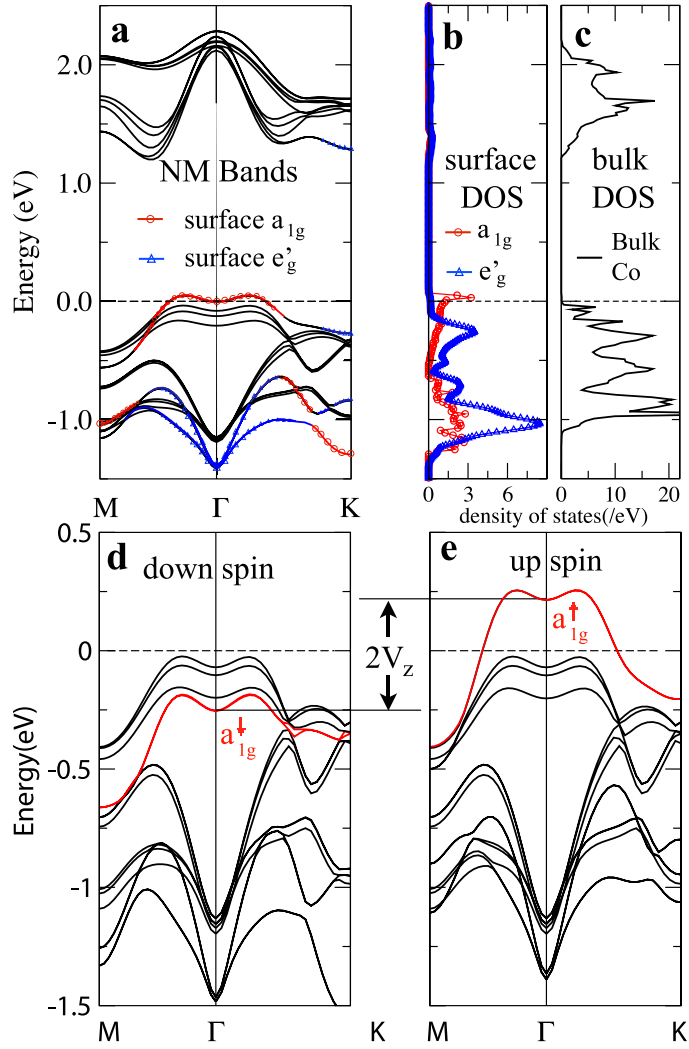}
\caption{(Color online) The calculated band structure and density of
  state (DOS) for NaCoO$_2$ slab with surface hole concentration
  $y$=0.3. (a) The band structure of non-magnetic (NM) state. The
  projections to the $a_{1g}$ and $e_g^\prime$ states of surface Co
  are indicated as red circle and blue triangle, respectively.  (b) and (c) The
  corresponding DOS for the surface and bulk Co sites,
  respectively. The character of insulating bulk with metallic surface
  is seen. (d) and (e) The spin-resolved band structure of
  ferromagnetic state with the $V_z$ defined as the exchange splitting
  of surface $a_{1g}$ state.}
\end{figure}

NaCoO$_2$ single crystal (with R\={3}m symmetry) is experimentally
available~\cite{crystal}, and it is a simple band insulator with band
gap more than 1.0eV~\cite{crystal}. Its bulk insulating behavior has
been indicated by transport~\cite{NaCoO-R} and NMR~\cite{NaCoO-NMR}
measurements. Electronically, the Co-$3d$ states splits into $t_{2g}$
and $e_g$ manifolds under oxygen octahedron crystal field, and all
$t_{2g}$ states are fully occupied (by 6 electrons) in the case
without Na deficiency (corresponding to nominal Co$^{3+}$ case),
resulting in an band gap between $t_{2g}$ and
$e_g$~\cite{crystal,NaCoO-NMR}. The layered crystal structure
guarantees that samples can be easily cleaved (Fig.1), and two kinds
of terminations, with or without top-most Na layer, may be
realized. Nevertheless, since the top-most Na$^{1+}$ ions are highly
mobile, its concentration can be tuned depending on experimental
conditions, resulting in surface hole doping (indicated as $y$) but
without modifying the surface structure (the CoO$_2$ layer)
significantly. In the extreme case, if all top-most Na are absent, 0.5
hole will be introduced (i.e. $y$=0.5).
We study the (001) surface of NaCoO$_2$ by using the first-principles
calculations based on the plane-wave ultra-soft pseudopotential
method, and the generalized gradient approximation (GGA) for the
exchange-correlation functional. A slab consisting of five NaCoO$_2$
unit layers thickness (Fig.1) and 20~\AA~vacuum region is used for the
surface study, and the SOC is included from the fully relativistic
pseudopotential. The cutoff energy for wave-function expansion is 30
Ry, and we use 12$\times$12 k-points mesh for the Brillouin zone
sampling. The calculations are well converged with respect to above
settings. The calculations are further supplemented by the
LDA+Gutzwiller method~\cite{LDAG-method}, in which the density
functional theory is combined with the Gutzwiller variational approach
such that the orbital fluctuation and kinetic renormalization coming
from correlation are all self-consistently treated.

By optimizing the surface (slab) structure without top-most Na layer,
our calculations confirm that the surface CoO$_2$ layer remain well
defined with only slightly (about 0.07\AA) modification of surface
Co-O bond length (see Fig.1). Another key result obtained from this
calculation is that the hole only goes into the surface CoO$_2$ layer,
while keeping all other CoO$_2$ layers inside the bulk
insulating. There is only one band cross the Fermi level (Fig.2),
which comes from one of the $t_{2g}$ states of the top-most Co sites,
while all other Co-$t_{2g}$ states inside the bulk are fully
occupied. It is therefore effectively a system with insulating bulk
but metallic surface, similar to topological
insulators~\cite{Kane_PRL_2005,Bernevig_PRL_2006,Zhang}. The key
difference lies in the fact that the surface state considered in this
work breaks the time reversal symmetry, in contrast to the case of
topological insulators. In reality, the surface hole doping can be
tuned by modifying the surface Na concentration, by interface charge
transfer, or simply by gating. In the case of proximity effect with a
superconductor, such surface doping can naturally arise since many
$s$-wave superconductors, like Al, Pb or Nb, are simple metals. In
such case, the insulating NaCoO$_2$ bulk can even serve as substrate
simultaneously. In the following discussions, we will therefore
neglect the top-most Na atoms and use the virtual crystal
approximation to simulate the surface doping effect.

\begin{figure}[tbp]
\includegraphics[clip,scale=0.25]{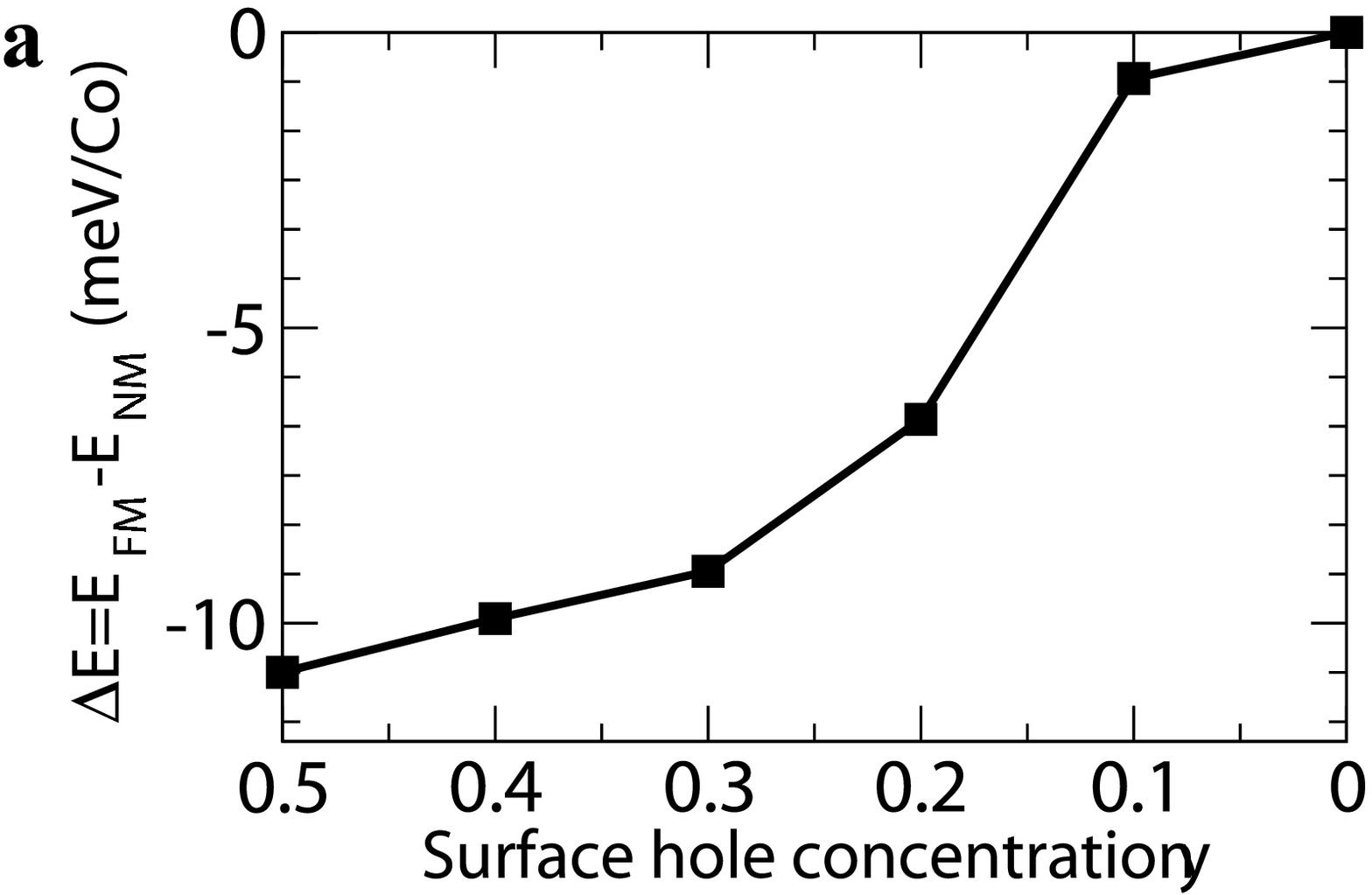} \ \ \ \ \
\includegraphics[clip,scale=0.25]{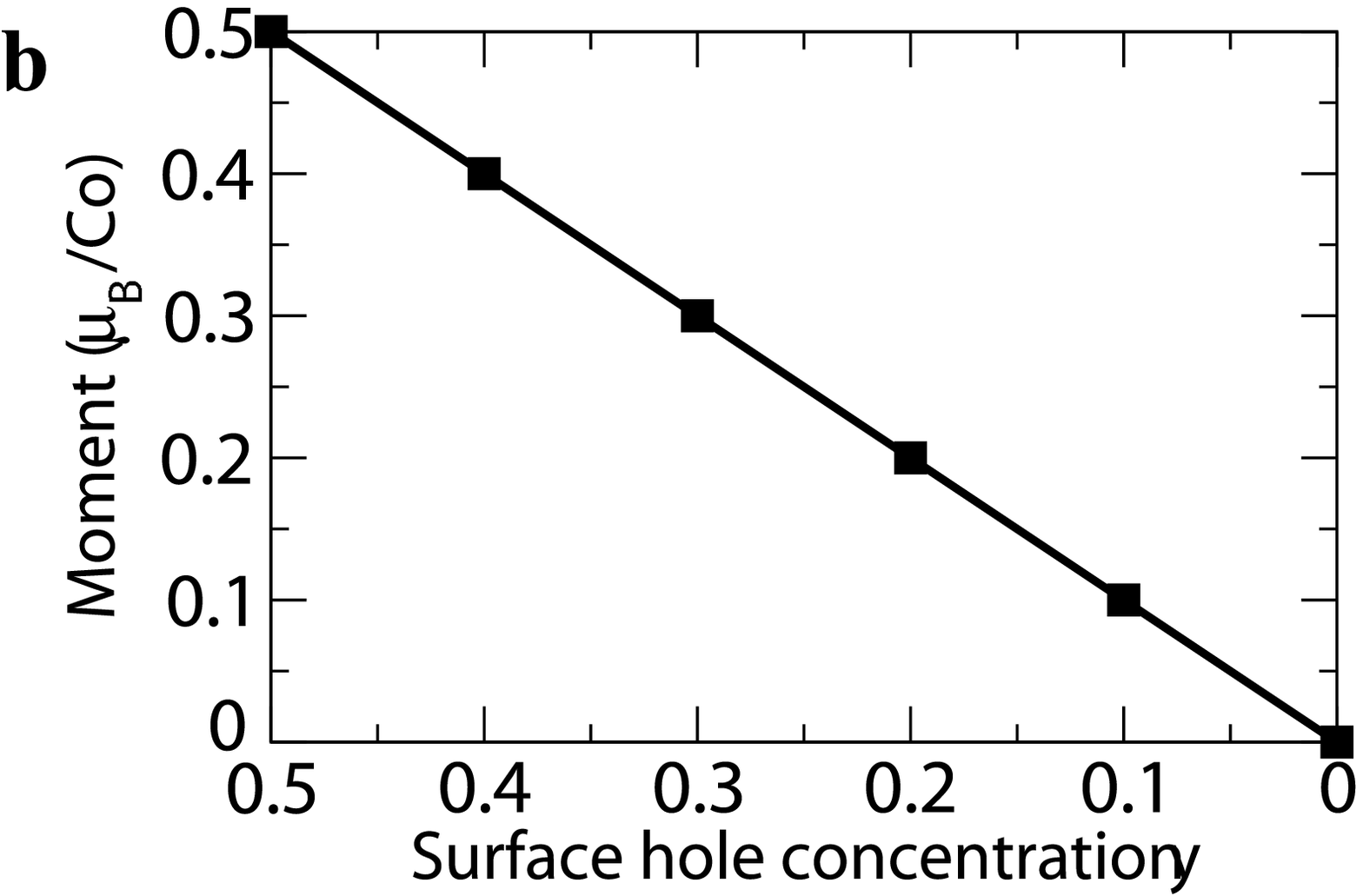}
\caption{(Color online) The calculated magnetic properties of
  NaCoO$_2$ surface as function of surface hole concentration $y$. (a)
  The stabilization energy of ferromagnetic (FM) state with respect to
  non-magnetic (NM) state. (b) The magnetic moment of surface CoO$_2$
  layer.}
\end{figure}

Fig.3 shows the magnetic properties of the NaCoO$_2$ surface as
function of hole concentration $y$. The spin polarization is
energetically favored for all hole concentration (0.5$<y<$0), in close
analogy to the layered AF phase of Na$_x$CoO$_2$ bulk
($x\sim$0.75)~\cite{NaCoO-phase,NaCoO-AF,LDAG}, where each CoO$_2$
layer (with $\sim$0.25 hole) orders ferromagnetically and the in-plane
ferromagnetism contributes mostly to the energy gain with relatively
weak inter-layer AF coupling. The stabilization of ferromagnetism at
NaCoO$_2$ surface can be intuitively understood from Stoner instability (also
similar to the discussion addressed for the layered AF bulk
Na$_{0.75}$CoO$_2$~\cite{LDAG}). Due to the elongation of oxygen
octahedra around Co sites along $c$-axis, the Co-$t_{2g}$ states will
further split into $a_{1g}$ and $e_g^\prime$ manifolds, with $a_{1g}$
higher in energy. The $a_{1g}$ state has mostly $3d_{3z^2-r^2}$
orbital character (with $z$ defined along $c$), whose in-plane dispersion
is relatively weak and ``M" shaped as observed in ARPES experiment~\cite{NaCoO-ARPES3}. 
As shown in Fig.2(a), the wide region
of nearly-flat-bands around the valence band maximum will produce a
sharp DOS peak near the Fermi level (see Fig.2(b)), which
leads to the Stoner instability, and favors the ferromagnetic (FM)
ground state for the surface. This mechanism is further supported by
the electronic structure of FM solution (Fig.2(c) and (d)), where only
the $a_{1g}$ state is strongly spin-polarized and the polarization of
$e_g^\prime$ states is small. In the FM state, the spin-polarization
is strong enough to make the top-most CoO$_2$ layer a half-metal, as
shown in Fig.2 and Fig.3, where the calculated magnetic moment
exactly equals to the number of holes. There is only one spin-channel
of $a_{1g}$ state crossing the Fermi level, resulting in a
half-metallic surface state with single sheet of Fermi surface.

We have to be aware of the effect of electron correlation beyond the
generalized gradient approximation (GGA) for the exchange-correlation
potential. The LDA+Gutzwiller method~\cite{LDAG-method} has been shown
to be a powerful tool to take into account the correlation effect, and
reproduce correctly the magnetic phase diagram of bulk
Na$_x$CoO$_2$~\cite{LDAG}. We have supplemented the LDA+Gutzwiller
calculations for the surface, and find that the surface ferromagnetism
is further stabilized (by about 8meV/surface Co for $y$=0.3). In fact,
in the study for the bulk Na$_{x}$CoO$_2$ ($x>$0.6), both GGA and
LDA+Gutzwiller give qualitatively the same result, well compared to
experiments~\cite{LDAG}. The AF state of bulk Na$_x$CoO$_2$ was
observed for $x>0.65$ with maximum $T_c$ around 25K for $x\sim$0.8,
the layered ordering with spin orientation perpendicular to the plane
was confirmed by neutron
experiments~\cite{NaCoO-neutron,NaCoO-neutron-2}, and its
half-metallicity of in-plane electronic structure is also supported by
the ARPES measurements~\cite{NaCoO-ARPES,NaCoO-ARPES-2}. Considering
the layered nature and the similarity between the bulk AF phase and
the surface, we conclude that a single band half metal can be realized
at NaCoO$_2$ surface with tunable hole concentration.

\begin{figure}[tbp]
\includegraphics[clip,scale=0.4]{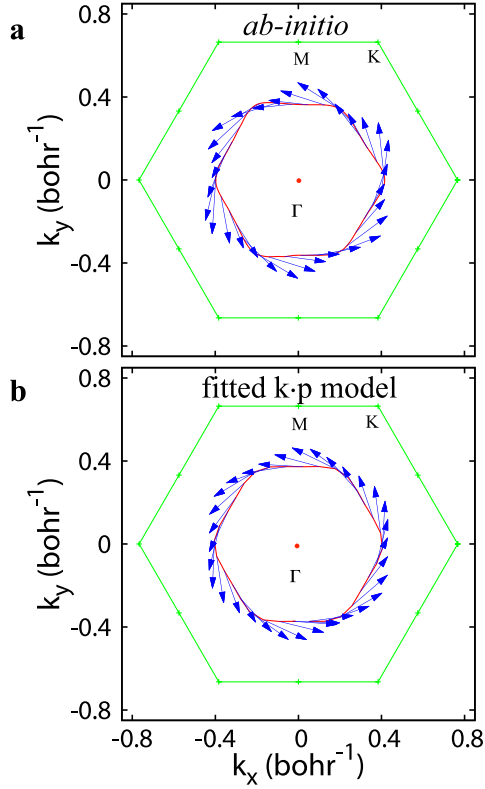}
\caption{(Color online) The Fermi surface and the helical spin texture
  of NaCoO$_2$ surface (for $y$=0.3) obtained from (a)
  first-principles calculations, and (b) evaluation of effective
  $k\cdot p$ model (Eq.1, see text part) with parameter $A$=0.28
  eV\AA$^2$, $B$=0.53 eV\AA$^6$, $C$=-2.4 eV\AA$^6$, $\alpha$=-0.066
  eV\AA, $V_z$=0.22eV. The in-plane components of spin are indicated
  as arrowed lines, while the perpendicular component (pointing to
  out-of-plane) is not shown.}
\end{figure}

Turning on the the SOC, the up and down spin bands will couple, while
the characteristic single sheet of Fermi surface still remains. In the
presence of a surface, the asymmetrical surface potential will produce
the Rashba type SOC (which is automatically included in the
first-principles calculations). It turns out that the Rashba SOC plays
important roles, resulting in the in-plane spin component, which has
helical spin texture for the states at Fermi level (see Fig.4).  The
in-plane component is actually rather strong, and contribute to more
than 10\% of the total moment from our first-principles
calculations. Considering the single $a_{1g}$ state of NaCoO$_2$
surface and the 3-fold rotation symmetry, an effective surface
Hamiltonian can be constructed as,
\begin{equation}
  H_0=\sum_{\bf k}[\xi({\bf k})+V_z\sigma_z+\alpha(k_y\sigma_{x}-k_x\sigma_{y})].
\end{equation}
where $\xi({\bf k})=Ak_{+}k_{-}+B(k_{+}^6+k_{-}^6)+C(k_{+}^3k_{-}^3)$
(with $k_\pm=k_x\pm ik_y$) gives the non-spin-polarized band
structure, $V_z\sigma_z$ is the exchange splitting, and the last term
is the Rashba type SOC due to the surface. Evaluating the eigen values
$\epsilon({\bf k})=\xi({\bf k})\pm\sqrt{V_z^2+\alpha^2{\bf k}^2}$ with
the parameters given in the caption of Fig.4, the Fermi surface and
its helical spin texture can be well reproduced (Fig.4(b)). Please
note the hole type carrier and the negative sign of Rashba coupling
$\alpha$ in our present case.

\begin{figure}[tbp]
\includegraphics[clip,scale=0.4]{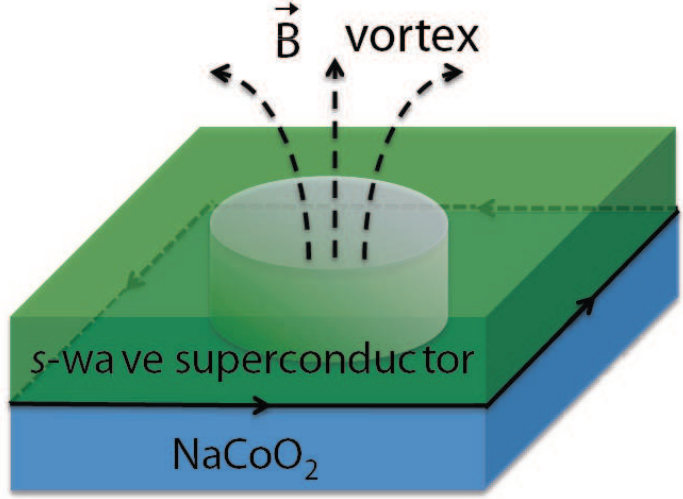}
\caption{(Color online) The schematic picture for the
  NaCoO$_2$/$s$-wave superconductor heterostructure where Majorana
  bound state is associated with the vortex core. The chiral Majorana
  edge state is also expect at the edge of interface.}
\end{figure}

When the NaCoO$_2$ surface comes into contact with an $s$-wave
superconductor, a pairing term $H_\Delta=\sum_{\bf k}[\Delta c_{{\bf
    k}\uparrow}^\dag c_{{\bf k}\downarrow}^\dag+h.c.]$ will be
generated by the proximity effect, and the full Hamiltonian reads
$H=H_0+H_\Delta$, which has been carefully studied
previously~\cite{Sarma,Alicea,Lee1,Lee2,HJ}. The dominant pairing
channel should have the spin polarized $p_x$+$ip_y$ symmetry, whose
order parameter is given as $\Delta_P^+({\bf k})=\frac{-\alpha
  k}{\sqrt{(V_z^2+\alpha^2k^2)}} \frac{k_y+ik_x}{k} \Delta$. In the
limit of large $V_z$, following Ref.~\cite{Alicea}, the
superconducting gap can be estimated as,

\begin{equation}
  E_g=\sqrt{\frac{2m^*\alpha^2}{V_z}(1+\frac{\mu}{V_z}) }\Delta\approx 0.22\Delta,
\end{equation}

which is sizable. Here we use the parameters, effective mass
$m^*$=5$m_0$, $\alpha=-0.066$eV\AA, and $V_z$=0.22eV, obtained from
our first principles calculations. The size of $\Delta$ is determined
by the superconducting gap of $s$-wave superconductor and its
interface coupling (hopping) with NaCoO$_2$ side.

%\section{Discussion}

Once the $p_x+ip_y$ superconductivity is realized through proximity
effect, the Majorana bound states associated with the vortex core will
be expected~\cite{proximity,Sarma,HJ}. The schematic experimental
setting for the detection of Majorana fermions is shown in Fig.5,
where the absence of magnetic insulating layer and/or magnetic field
is the essential difference compared to earlier
proposals~\cite{proximity,Sarma,Alicea}. In additional, due to the
broken time reversal symmetry (different from Fu and Kane's
proposal~\cite{proximity}), a chiral Majorana edge state should exist
at the interface (Fig.5).  Since the single band half-metallic
character of NaCoO$_2$ surface can be realized in a large energy
window (of about 0.2eV), the fine tuning of chemical potential is not
necessary.  The large Fermi surface size and the high carrier density
should be also helpful for sizable proximity effect, against the
possible localization due to disorders. Among several possible choices
of the superconductor, it is particularly interesting to consider the
Na$_x$CoO$_2$ superconductor with $x$ around
$0.35$~\cite{NaCoO-phase,NaCoO-SC}, since the lattice structures are
well matched. The NaCoO$_2$/$s$-wave superconductor heterostructure,
therefore, provide us a new and simple platform for realizing
topological superconductivity and Majorana fermion bound states.  This
work can be generalized in several directions. Similar predictions can
be also made for AF Na$_x$CoO$_2$ ($x>$0.65) thin film with odd number
of layers (which contribute to odd number of Fermi surfaces), or for
LiCoO$_2$, which has the same crystal and electronic structure as
NaCoO$_2$~\cite{LiCoO}.

%\section{Methods}

%\section{Acknowledgement}

We acknowledge the supports from NSF of China and that from the 973
program of China (No.2007CB925000, No.2010CB923000), the US NSF under
grant numbers DMR-0904264, and the Keck Foundation.

\end{document}